\documentclass[aps,prl,reprint,groupedaddress,showpacs]{revtex4-1}
\usepackage{graphicx}
\usepackage{bm}
\usepackage{amsmath}
\usepackage{psfrag}

\begin{document}

\title{Nonlinear degradation enhanced transport of morphogens performing subdiffusion}
\author{Sergei Fedotov}
\author{Steven Falconer}
\affiliation{School of Mathematics, The University of Manchester, Manchester
M60 1QD, United Kingdom}
\date{\today }

\begin{abstract}
We study a morphogen gradient formation under nonlinear degradation and subdiffusive transport.
In the long time limit we obtain the nonlinear effect of degradation enhanced diffusion, resulting from the interaction of non-Markovian subdiffusive transport with a nonlinear reaction.
We find the stationary profile of power-law type, which has implications for robustness, with the shape of the profile being controlled by the anomalous exponent.
Far away from the source of morphogens, any changes in rate of production are not felt.
\end{abstract}

\maketitle

\section{Introduction}

During the development of an organism, a key stage is the differentiation of cell types \cite{Rogers2011}.
It is known that the differentiation of these identical cells, into different and
distinct cell types is controlled by a signalling molecule called a
morphogen \cite{Entchev2000}.
One of the most widely studied organisms in the field of morphogenesis is the \textit{Drosophila}, common fruit fly, and particularly the development of its wings.
The wings begin as a multinucleated mass of identical cells within a membrane, in the early embryo, called an imaginal disc.
A morphogen from the TGF-$\beta $ superfamily called decapentaplegic (Dpp) is secreted by a narrow strip of cells, from which it diffuses in essentially one dimension and degrades, causing a concentration gradient to form.
The production, diffusion, and degradation of morphogens are controlled by a complex set of positive and negative feedback loops \cite{Barkai2009}.
The cells in the imaginal disc react to the concentration gradient at discrete levels \cite{Ashe2006}, enabling them to determine their position within the disc.
From knowing their position, the cells are able to differentiate themselves to carry out different functions within the developed wing.
Thus, to prevent mutations it is essential that the concentration gradient built up is robust to fluctuations in secretion rate due to genetic alterations, temperature changes, or any other environmental effects \cite{Eldar2002}.

There are differing thoughts on the mechanism behind the diffusion of the morphogen, whether the transport is primarily extracellular or intracellular \cite{Bressloff2013}.
Whether it is able to diffuse freely through the, essentially, 2D plane of the imaginal disc or; whether the molecules are passed over between neighbouring cells, a process named transcytosis \cite{Bollenbach2005}.
It is thought that some morphogens require intracellular trafficking, whilst others may diffuse freely \cite{Kicheva2007}.
However, regardless of the specific mechanism, it is known that morphogens do form long range concentration gradients, and that the robustness of the concentration gradient is of the utmost importance \cite{Bollenbach2005, Barkai2009,Rogers2011}.

The standard model for morphogen transport is the diffusion equation with degradation term
\begin{equation}
\frac{\partial \rho }{\partial t}=D\frac{\partial ^{2}\rho }{\partial x^{2}} -\theta \rho ,  \label{main}
\end{equation}
where $\rho (x,t)$ is the density of morphogen, $D$ is the diffusion coefficient, $\theta $ is the degradation rate.
This equation together with the boundary condition with the constant source term at $x=0$ gives a stationary concentration distribution which decays exponentially.
It has been argued that an exponential profile cannot be robust to fluctuations in environmental conditions and production rate \cite{Eldar2003}.
Therefore, the aforementioned authors argued a power law profile is preferable.
Experiments have shown that in some circumstances, a power law decay is observed for the morphogen profile \cite{Han2005}.
One way to obtain this profile is to assume that the morphogens must decay rapidly close to their source, whilst decaying at a much slower rate over the rest of the area.
In other words, the degradation rate is an increasing function of the local concentration of diffusing morphogens.
In this case the only modification to (\ref{main}) is the nonlinear rate $\theta (\rho )$.
The topic has been tackled in \cite{Eldar2002,Eldar2003}, where the authors dubbed this the `self enhanced degradation' of morphogens.

The robust stationary profile can be found from
\begin{equation}
D\frac{d^{2}\rho _{st}(x)}{dx^{2}}=k\rho _{st}^{2}(x),
\end{equation}
with the boundary condition at $x=0:$  $-Dd\rho _{st}/dx=g$.
This leads to algebraic decay in the tails of the spatial distribution,
\begin{equation}\label{eq:std_tails}
\rho _{st}(x)\sim \frac{A}{x^{2}}, \quad x\to \infty,
\end{equation}
where the amplitude $A$ is independent of production term $g$.
In \cite{Eldar2003} the authors take the independence of production rate from amplitude of the profile to be a key indicator of the robustness of the profile.
A nonlinear degradation rate can arise from the situation in which the morphogen increases the production of a molecule which in turn increases the rate of morphogen degradation.
In the example of the Drosophila fly, the morphogen Shh is responsible for the expression of a receptor which both transduces the Shh signal, and mediates the degradation of the morphogen \cite{Chen1996,Gordon2011a,*Muratov2011}.

Hornung, Berkowitz, Barkai \cite{Hornung} published the first paper in which subdiffusion of morphogens considered.
Subdiffusion is an observed natural phenomenon, seen in the diffusion of proteins in the cytoplasm and the nucleus of eukaryotic cells \cite{Weiss2004,Golding2006}, along the surface of a cell membrane \cite{Saxton2001,Min2005}, and has been suggested to explain morphogen movement in a heterogeneous environment of HSPG proteins.
For the anomalous subdiffusion the mean squared displacement grows sub-linearly with time $\langle x^{2}(t)\rangle \sim t^{\mu }$, where $\mu <1$ is the anomalous exponent.
Following on from Hornung \textit{et. al.} \cite{Hornung}, several attempts have been made to take into account subdiffusion for the analysis of morphogen gradient formation \cite{Kruse2008,Yuste2010,Boon2012,Fedotov2013}.
Kruse and Iomin \cite{Kruse2008} developed a  microscopic model of the receptor mediated transport in a subdiffusive medium.
They found subdiffusive and  superdiffusive spreading of morphogens.
Yuste \textit{et. al } \cite{Yuste2010} analyzed the gradient formation of subdiffusive morphogens by using the reaction-subdiffusion equation obtained from a classical continuous time random walk (CTRW)
\begin{equation}
\frac{\partial \rho }{\partial t}=D_{\mu }\frac{\partial ^{2}}{\partial x^{2}}\left[ e^{-\theta (x)t}\mathcal{D}_{t}^{1-\mu }\left[ e^{\theta (x)t}\rho(x,t)\right] \right] -\theta (x)\rho ,  \label{main2}
\end{equation}
where $D_{\mu }$ is the fractional diffusion coefficient, and $\mathcal{D}_{t}^{1-\mu }$ represents the Riemann-Liouville fractional derivative of order $1-\mu$
\begin{equation}
\mathcal{D}_{t}^{1-\mu }f(x,t)=\frac{1}{\Gamma (\mu )}\frac{\partial }{\partial t}\int_{0}^{t}\frac{f(x,t^{\prime })}{\left( t-t^{\prime}\right) ^{1-\mu }}dt^{\prime }.  \label{RL}
\end{equation}
The main difference of this work to that of \cite{Hornung} is that here the particles are not protected during trapping events.
A stationary profile does not exist in the model of \cite{Hornung}; the authors obtained only a non-stationary exponential profile in space, with a power law decay of amplitude in time.
Yuste, Abad, and Lindenberg \cite{Yuste2010} found the stationary exponential profile and analyzed the interaction of subdiffusion and space-dependent degradation.
A diffusion equation with a power law density dependent diffusion coefficient and nonlinear degradation has been analyzed in the recent paper \cite{Boon2012}.
The modified fractional Fokker-Planck equation was used for the analysis of morphogen gradient formation in \cite{Fedotov2013}, where they employed the random death process in a such way that the degradation term acts like a tempering of the waiting time distribution.
This leads to the unusual effect of the dependence of the diffusion coefficient on the degradation rate.
The authors considered only a linear death process and did not consider feedback effects in the degradation rate, and indeed many current models do not either.

The main purpose of this work is to analyze the interaction of the nonlinear degradation with non-Markovian subdiffusion, and its implications on the stationary structure.
The result of this interaction is \textit{degradation enhanced diffusion} in the long time limit.
The gradient profile can be found from the nonlinear stationary equation for which \textit{the diffusion coefficient is a nonlinear function of the nonlinear reaction rate}
\begin{equation}\label{eq:main_result}
\frac{d^{2}}{dx^{2}}\left( D_{\theta }(\rho _{st}(x))\rho _{st}(x)\right) = \theta (\rho _{st}(x))\rho _{st}(x).
\end{equation}
Here the diffusion coefficient $D_{\theta }$ is
\begin{equation}\label{eq:nonlinear_diffn}
D_{\theta }(\rho _{st}(x))=\frac{a^{2}\left[ \theta (\rho _{st}(x))\right]^{1-\mu (x)}}{2\tau _{0}{}^{\mu (x)}},
\end{equation}
and $ \tau_0 $ is the time parameter, and $ \mu(x) $ the space dependent anomalous exponent.
This unusual form of nonlinear diffusion coefficient is a result of the interaction between non-Markovian transport and nonlinearity.
The interaction leads directly to a \textit{degradation enhanced diffusion}.
This effect does not exist for the Markovian random walk model presented in \cite{Boon2012}.
We also would like to direct the reader to the interesting paper on the influence of coupling between diffusion and degradation on the morphogen gradient formation \cite{Anatoly}.
\section{Subdiffusive Transport and Nonlinear Degradation}
We describe a random morphogen molecule's movement in an extracellular surrounding as follows.
We assume that molecules are produced at the boundary $x=0$ of the semi-infinite domain $[0,\infty )$ at the given constant rate $g $, and perform the classical continuous-time random walk involving symmetrical random jumps of length $a$ with random waiting time $T_{x}$ between jumps.
If we assume that this random time is exponentially distributed with the rate parameter $\lambda $ then on the macroscopic level we obtain the classical diffusion term in (\ref{main}) with diffusion coefficient $D=\lambda a^{2}/2$.
In this paper we consider the subdiffusive behaviour for morphogen molecules when the residence time $T_{x}$ has the survival probability $\Psi(x,t)=\Pr \left[ T_{x}>t\right] $ given by the Mittag-Leffler function \cite{Hilfer}
\begin{equation}
\Psi (x,t)=E_{\mu (x)}\left[ -\left( \frac{t}{\tau _{0}}\right) ^{\mu (x)}\right] ,\quad 0<\mu (x)<1.
\end{equation}
The Mittag-Leffler distribution is characterized by its interpolation between short time stretched exponential, and long time power law asymptotics
\begin{equation}\label{eq:Mittag_asymp}
\Psi (x,t)\simeq
\begin{cases}
\frac{1}{\Gamma (1+\mu (x))} e^{-\left( \frac{t}{\tau _{0}}\right) ^{-\mu (x)}}, & t<<1, \\
\frac{1}{\Gamma (1-\mu (x))} (\frac{t}{\tau _{0}})^{-\mu (x)},\quad & t\to \infty .
\end{cases}
\end{equation}
This distribution leads to the divergence of the mean waiting time
\begin{equation}
\bar{T}_{x} = - \int_{0}^{\infty } t\frac{\partial \Psi (x,t)}{\partial t}dt ,\quad 0<\mu (x)<1,
\end{equation}
which explains the slow subdiffusive behaviour.
This emerges from the CTRW scheme when a molecule becomes immobilized in a region of space, and the mean escape time diverges.
The reasons for trapping are many, and vary on the circumstance.
The particles could be trapped in intracellular space whilst cell surface receptors are occupied \cite{Kruse2008,Barkai2009}.
It could be that a particle enters a region with a very complicated geometry, such a dendritic spine, and struggles to escape \cite{Santamaria2011}.
It could be immobilized by some chemical reactions.

We describe the morphogen degradation by the mass action law involving the nonlinear reaction term
\begin{equation}
\theta (\rho )\rho ,
\end{equation}
where the reaction rate $\theta (\rho )$ depends on the mean density $\rho $.
The importance of a nonlinear reaction rate lies in the effect of self-enhanced ligand degradation which underlies the robustness of morphogen gradients \cite{Eldar2002,Eldar2003} (see also \cite{Gordon2011a}).
It should be noted that the authors of \cite{Hornung} consider very different model in which morphogen molecules are protected during the trapping time $T_{x}$ and degradation occurs instantaneously at the end of a waiting time with given probability.

Our assumptions lead to the following nonlinear reaction-subdiffusion equation for the mean density of morphogen molecules \cite{Yadav2006}
\begin{align}
\frac{\partial \rho }{\partial t} = \frac{\partial ^{2}}{\partial x^{2}} &\left[  D_{\mu }(x)e^{-\int_{0}^{t}\theta (\rho )ds} \mathcal{D}_{t}^{1-\mu (x)}\left[e^{\int_{0}^{t}\theta (\rho )ds}\rho (x,t)\right] \right] \nonumber \\
 & \qquad -\theta(\rho)\rho ,  \label{main3}
\end{align}
where
\begin{equation}
D_{\mu }(x)=\frac{a^{2}}{2\tau _{0}^{\mu (x)}},
\end{equation}
$ a $ is the jump length and $ \tau_0 $ is the time parameter.
See also \cite{mendez2010reaction} pp.48-52.
The main characteristic of this reaction-transport equation is that the reaction and transport are not additive.
Due to the non-Markovian nature of subdiffusion, it is not possible to separate reaction as an extra term on the RHS as is the case for a regular diffusion like \eqref{main}.
Instead, reaction terms also appear mixed in the derivative term as an exponential factor, as seen above.
The presence of the Riemann-Liouville derivative indicates a long memory in the process, presenting itself in the integral over time, making it strongly non-Markovian.

It turns out that in the long time limit this equation leads to a nonlinear diffusion with a diffusion coefficient depending on the nonlinear degradation.
Note that nonlinear diffusion has been analyzed in \cite{Boon2012}, where the authors introduced a nonlinear dependence of diffusion coefficient of the density independently from the reaction.
Moreover, this nonlinear diffusion is independent from degradation.
In this paper we show how nonlinear diffusion emerges naturally from the microscopic random walk for which the nonlinear diffusion and degradation are not independent.
We also take into account a spatially non-uniform distribution of anomalous exponent $\mu (x)$.
We have shown previously that any spatial variation in the anomalous exponent $\mu $ leads to a drastic change in the stationary behavior of the fractional subdiffusive equations \cite{Fedotov2012}, a phenomenon called anomalous aggregation \cite{Fedotov2011}.
Note that the robustness of the stationary profile of diffusing morphogens is the most important feature \cite{Eldar2002}.

The fractional reaction-transport equation (\ref{main3}) can be rewritten in the compact form
\begin{equation}\label{eq:RsD_i}
\frac{\partial \rho }{\partial t}=\frac{a^{2}}{2}\frac{\partial ^{2}i(x,t)}{\partial x^{2}}-\theta (\rho )\rho(x,t),
\end{equation}
where $i(x,t)$ is the total escape rate from the point $x$.
It follows from equation \eqref{main3} that it can be written as
\begin{equation}
i(x,t)=\frac{e^{-\int_{0}^{t}\theta (\rho )ds}}{\tau _{0}{}^{\mu (x)}}\mathcal{D}_{t}^{1-\mu (x)}\left[ e^{\int_{0}^{t}\theta (\rho )ds}\rho (x,t)\right].  \label{eq:fractional_i}
\end{equation}
Different choices for the form of the escape rate can lead to many interesting equations in the diffusion limit \cite{fedotov2013nonlinear}.
\section{Stationary Morphogen Profile}
\subsection{Linear Degradation}
In a previous publication \cite{Fedotov2013}, we gave full details on how the linear version of reaction-subdiffusion equation \eqref{main3} approaches a stationary diffusion.
In this section, we will re-cap this, and extend to the current nonlinear consideration.
The linear reaction-subdiffusion equation considered in \cite{Fedotov2013} differs from \eqref{eq:RsD_i} with the total escape rate $ i $ being given by:
\begin{equation}
i(x,t)=\frac{e^{-\theta(x) t}}{\tau _{0}{}^{\mu (x)}}\mathcal{D}_{t}^{1-\mu(x)}\left[ e^{\theta(x) t}\rho (x,t)\right] .  \label{eq:fractional_i_linear}
\end{equation}

To obtain a stationary solution for the system, it is necessary to introduce a flux of new particles, $g$.
We choose to implement this on the boundary $x=0$, this directly corresponds to the morphogen problem, where particles are produced from a point source.
For conservation reasons, the logical choice for production rate is $ g = \int_{0}^{\infty }\theta(x) \rho _{st}(x)dx $ \cite{Yuste2010}.

The Laplace transform of the integral escape rate \eqref{eq:fractional_i_linear} is found by the shift theorem:
\begin{equation}
\hat{i}(x,s) = \int_{0}^{\infty} i(x,t) e^{- s t} dt = \frac{[s + \theta(x)]^{1 - \mu(x)}}{\tau _{0}{}^{\mu (x)}} \hat{\rho}(x,s).
\end{equation}
The limit $t \to \infty $ corresponds to the limit $s \to 0 $ of the Laplace variable.
We write for the stationary total escape rate $i_{st} (x) $
\begin{equation}  \label{eq:stationary_i_linear}
	i_{st}(x) = \lim_{s \to 0} s \hat{i}(x,s) =\frac{\theta(x)^{1 - \mu(x)}}{\tau _{0}{}^{\mu (x)}} \rho_{st}(x),
\end{equation}
where $ \rho_{st}(x) = \lim_{s \to 0} s \hat{\rho}(x,s) $.
This follows from the standard final value theorem stating that when $ \lim_{t \to \infty} f(t)$ exists, then $\lim_{t \to \infty} f(t) = \lim_{s \to 0} s \hat{f}(s)$.
Note that the equation \eqref{eq:stationary_i_linear} has a Markovian form, since the escape rate can be written in the form $ i_{st}(x) = \lambda \rho_{st}(x)$, where $\lambda = \theta(x)^{1 - \mu(x)}/ \tau _{0}{}^{\mu (x)}$ now depends upon the degradation rate.
This shows the transition from subdiffusive dynamics, to asymptotically normal diffusion.

Consider for contrast that if the death rate is constant in time and space, and independent of $\rho $, and the drift is zero; then we find an analytic result for the stationary gradient distribution, as an exponential function \cite{Yuste2010}.
The stationary profile is given by
\begin{equation}\label{eq:exponential}
\rho _{st}(x)=\frac{g}{\sqrt{\theta^{2-\mu} D_{\mu }}}\exp \left[ -\sqrt{\frac{\theta^{\mu} }{D_{\mu }}}x\right],
\end{equation}
and, as mentioned, the full details can be found in \cite{Fedotov2013}.
\subsection{Nonlinear Degradation}
It has been argued that even for subdiffusion a stationary exponential morphogen profile cannot be robust to fluctuations in both environmental effects and production rate \cite{Eldar2002,Eldar2003}.
The purpose of this subsection is to show that a robust stationary morphogen profile can be found as a result of the interaction of non-Markovian subdiffusion and nonlinear degradation.
The question now is how to take into account a nonlinear reaction term.
Actually, it turns out that the same techniques can be used as for the previous linear case.
From the total escape rate \eqref{eq:fractional_i} we seek to use the Laplace transform shift theorem, and the Tauberian theorem, to find the stationary behavior.
If the stationary distribution exists then,
\begin{equation}
\lim_{t \to \infty} \frac{1}{t} \int_{0}^{t} \theta(\rho(x,s)) ds= \theta(\rho _{st}(x)).
\end{equation}
As a result, $ e^{ - \int_{0}^{t} \theta(\rho(x,s)) ds } \to e^{ - \theta( \rho_{st}(x) ) t } $ as $ t \to \infty $.
This argument makes the shift theorem directly applicable, leading to the stationary escape rate for the nonlinear case,
\begin{equation}
i_{st}(x)=\frac{ [\theta(\rho _{st}(x))]^{1-\mu (x)}}{\tau _{0}^{\mu (x)}} \rho_{st}(x).
\end{equation}
Note that similar arguments have been made in \cite{Froemberg2008}.
For this escape rate, equation \eqref{eq:RsD_i} can be rewritten in a stationary form as
\begin{equation}
\frac{a^{2}}{2}\frac{d^{2}i_{st}(x)}{dx^{2}}=\theta (\rho _{st}(x))\rho_{st}(x).
\end{equation}
Finally the stationary nonlinear reaction-subdiffusion equation takes the form of a nonlinear second order ODE
\begin{equation}\label{eq:stationary_main}
\frac{d^{2}}{dx^{2}}\left( \frac{a^{2}\left[ \theta (\rho _{st}(x))\right]^{1-\mu (x)}}{2\tau _{0}{}^{\mu (x)}}\rho _{st}(x)\right) =\theta (\rho_{st}(x))\rho _{st}(x).
\end{equation}
This equation has the form of equation \eqref{eq:main_result} where the diffusion coefficient $ D_\theta $ is an increasing function of the nonlinear reaction rate \eqref{eq:nonlinear_diffn}.

Let us consider the commonly studied case of an n-fold superlinear reaction term in the stationary nonlinear reaction-subdiffusion equation \eqref{eq:stationary_main}, corresponding to a reaction term
\begin{equation}
\theta (\rho)=k\rho^{n-1},
\end{equation}
where $ k $ is the reaction constant.
In what follows, we consider only $ \mu(x) = \mu = const $.
Here the total escape rate is given by
\begin{equation}
i(x,t)=\frac{e^{-k \int_{0}^{t} \rho^{n-1} ds}}{\tau _{0}{}^{\mu }}\mathcal{D}_{t}^{1-\mu }\left[ e^{k \int_{0}^{t} \rho^{n-1} ds}\rho (x,t) \right] .
\label{eq:fractional_i_nonlinear}
\end{equation}
We can write the nonlinear equation \eqref{eq:stationary_main} as
\begin{equation}\label{eq:stationary_ode}
 D_{\mu }k^{1-\mu }\frac{d^{2}}{dx^{2}} \left[ (\rho _{st}(x))^{ \left( n-1 \right) \left( 1-\mu \right) +1} \right] = k\rho _{st}^{n}(x),
\end{equation}
where
\begin{equation}
D_{\mu }=\frac{a^{2}}{2\tau _{0}{}^{\mu }}.
\end{equation}
The boundary conditions are given by
\begin{equation}
- D_{\mu }k^{1-\mu } \left. \frac{d}{d x} \left[ (\rho _{st}(x))^{ \left( n-1 \right) \left( 1-\mu \right) +1} \right] \right|_{x=0} = g,
\end{equation}
at $ x = 0 $, and $ \lim_{x \to \infty} \rho_{st}(x) = 0 $.

\begin{figure}[tbp]
\includegraphics[scale=0.53]{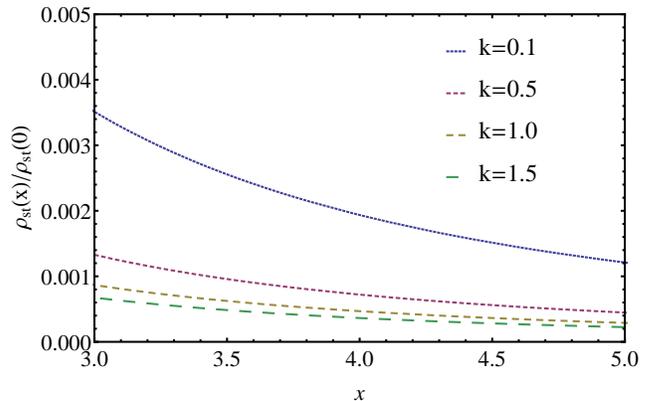}
\caption{Stationary profile \eqref{eq:Soln_Boon} with parameters: $\protect n=2, \mu =0.9$, $\protect\tau _{0}=0.001$, $g=10$, $ a = 0.01 $.}
\label{fig:PowLaw}
\end{figure}

Equation \eqref{eq:stationary_main} is written in the form of a balance equation between reaction and transport, however for a reaction-subdiffusion equation the two cannot be separated.
The right hand side of equation \eqref{eq:stationary_main} is a pure reaction, balanced with the mixed reaction-transport on the other side.
We can make the interesting observation that if we multiply both sides of the equation by $ k^{\mu-1} $, then we obtain exactly the same form for the equation as from the nonlinear theory \cite{Boon2012}.
In their model, the nonlinear diffusion is completely separate from the reaction.
The authors introduced two nonlinear functions $F$ \& $G$ into their Markovian random walk model.
However, in our non-Markovian model the nonlinear diffusion and the reaction are not independent and cannot be separated.
We showed that the assumption of a nonlinear reaction leads directly to a `degradation enhanced diffusion'.
This comes about from the nontrivial interaction between subdiffusion and reaction, which is a result of the long range memory of the underlying random walk model.
In regular diffusion, such as in the model \cite{Boon2012}, a stationary profile can be obtained by simply equating the time derivative to zero; in subdiffusion that is not the case (see equation \eqref{main3}).
Note that here the nonlinear diffusion dependence on reaction rate is not postulated, but emerges naturally from the interaction of subdiffusion and nonlinear reactions.
Despite the essential differences between the non-Markovian equation \eqref{main3} and that which was presented in \cite{Boon2012}, the stationary equations \eqref{eq:stationary_ode} are similar and can be solved in the same way
\begin{equation}\label{eq:Soln_Boon}
\rho _{st}(x)=\rho _{st}(0)\left( 1+\frac{x}{x_{0}}\right) ^{-\frac{2}{\mu(n-1)}},
\end{equation}
where
\begin{align}
\rho _{st}(0)& =\left( g^{\ast }\sqrt{\frac{\alpha +n}{2\alpha }}\right) ^{\frac{2}{\alpha +n}},\quad g^{\ast }=\frac{g}{\sqrt{D_{\mu }k^{2-\mu }}}, \nonumber \\
x_{0}& =\frac{2\alpha }{n-\alpha }\left( {g^{\ast }}\right) ^{-\frac{n-\alpha }{\alpha +n}}\left( \frac{\alpha +n}{2\alpha }\right) ^{\frac{\alpha }{\alpha +n}}\sqrt{\frac{D_{\mu }}{k^{\mu }}}, \nonumber \\
\alpha & =(n-1)(1-\mu )+1.
\end{align}
When $x/x_0 \gg 1$  we obtain the power law profile:
\begin{equation}\label{eq:sub_tails}
	\rho _{st}(x)\sim \frac{A}{ x^{\frac{2}{\mu(n-1)}}},\quad x\to \infty,
\end{equation}
where the amplitude 
\begin{equation}\label{eq:A}
	A = \rho_{st}(0) x_{0}^{\frac{2}{\mu(n-1)}},
\end{equation}
is independent of the morphogen production rate $ g $.
In the tails, this profile has an inverse dependence on the constant degradation rate $ k $, as illustrated in FIG.~\ref{fig:PowLaw}.
The effect of decreasing $ \mu $ is a decrease in the amplitude of the tails.
This should be expected since the interpretation of $ \mu $ is as a parameter controlling the strength of spatial trapping of particles, with decreasing $ \mu $ increasing trapping strength, as seen in the behaviour of the survival function \eqref{eq:Mittag_asymp}.
To counteract the trapping, the rate of diffusion is increased by the degradation rate, which we term \textit{degradation enhanced diffusion}.
Comparing tail behavior in the standard diffusion \eqref{eq:std_tails} with that of subdiffusion \eqref{eq:sub_tails}, the impact of $ \mu $ is clear.
\subsection{Robustness}
Let us now discuss the robustness of the profile \eqref{eq:Soln_Boon} with respect to the morphogen production rate $ g $.
It is convenient to write \eqref{eq:Soln_Boon} in the following way:
\begin{equation}\label{eq:stationary_robust}
	\rho_{st}(x) = \frac{A}{(x_{0} + x)^{\frac{2}{\mu(n-1)}}},
\end{equation}
where $ A $ is defined in \eqref{eq:A}.
The only parameter in \eqref{eq:stationary_robust} which is dependent on $ g $ is $ x_{0} $:
\begin{equation}\label{eq:robust_x0}
	x_{0} = \frac{B}{g^{\frac{n-\alpha}{n+\alpha}}},
\end{equation}
where the parameter $ B $ is independent of $ g $, and $ \frac{n-\alpha}{n+\alpha} > 0 $.
From \eqref{eq:stationary_robust} and \eqref{eq:robust_x0} it is clear that a change in $ g $ produces a uniform shift in the stationary profile along the $ x $-axis.

\begin{figure}[h]
	\includegraphics[scale=0.53]{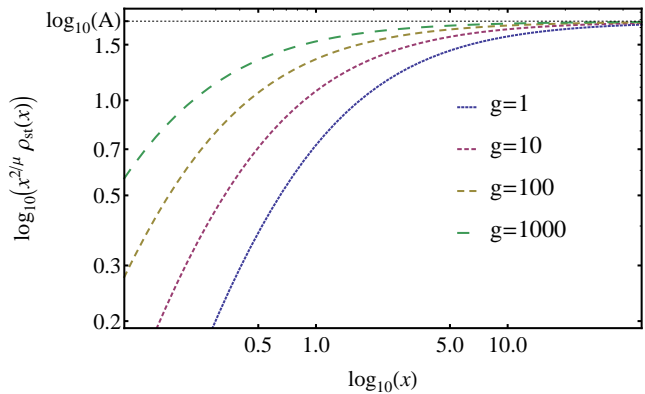}
	\caption{$\log_{10}(x^{2/\mu}\rho_{st}(x))$ against $\log_{10}(x)$ illustrating of convergence to the robust profile $\rho_{st}(x) \sim x^{2/\mu}$.
	Effect of varying production rate $ g $ for Eq.~\eqref{eq:Soln_Boon} with parameters: $\protect n=2, \mu =0.9$, $\protect\tau _{0}=0.001$, $ k=1 $, $ a = 0.01 $.
	The parameter $A$ is defined in \eqref{eq:sub_tails}.}
	\label{fig:loglog_g}
\end{figure}

The robustness of the profile \eqref{eq:Soln_Boon} to changes in the morphogen production rate $ g $ can be assessed with a standard sensitivity analysis involving the relation
\begin{equation}
	\delta \rho_{st}(x) = \frac{\partial \rho_{st}(x)}{\partial g} \delta g,
\end{equation}
where $ \delta g $ is a small change in the production rate.
The non-dimensional robustness parameter $ R $ can be introduced in several ways (see, for example \cite{Yuste2010, Eldar2003b, Bollenbach2005}).
We choose to define this measure from the following relation:
\begin{equation}
	\frac{\delta \rho_{st}(x)}{\rho_{st}(x)} = \frac{1}{R} \frac{\delta g}{g},
\end{equation}
where 
\begin{equation}
 \frac{1}{R} = \frac{g}{\rho_{st}(x)} \frac{\partial \rho_{st}(x)}{\partial x} \frac{\partial x_{0}}{\partial g}.
\end{equation}
This relates the relative change in the density at a given point $ x $ with respect to the relative change in $ g $.
For large values of $ R $ the system is robust.
For the profile given by \eqref{eq:stationary_robust}, we find the expression for $ R $ to be
\begin{equation}\label{eq:R_explicit}
	R = \frac{n+\alpha}{n-\alpha} \frac{\lambda}{x_{0}},
\end{equation}
where $ \lambda $ is the local spatial decay length defined as:
\begin{equation}\label{eq:lambda}
	\lambda = - \frac{\rho_{st}(x)}{\frac{\partial }{\partial x}\rho_{st}(x)} = \frac{\mu}{2} (n-1)(x_{0} + x).
\end{equation}
Notice that this expression for the decay length \eqref{eq:lambda} depends explicitly on the anomalous exponent $ \mu $.
The exponential profile \eqref{eq:exponential} has a corresponding value of $ R = 1 $ and it is not robust to changes in $ g $.
From \eqref{eq:robust_x0}, \eqref{eq:R_explicit}, and \eqref{eq:lambda}, it is clear that $ R \to \infty $ as either of the parameters $ g \to \infty $, $ x \to \infty$.
This indicates power-law profile \eqref{eq:Soln_Boon} is robust to changes in the production rate $ g $.
This can also be seen in FIG.~\ref{fig:loglog_g}, where increasing values of the production rate causes convergence to the robust power-law profile even for smaller values of $ x $.
Additionally, for the values of $\log_{10}(x)>10$ we have almost complete convergence.
As mentioned by previous authors \cite{Eldar2003}, this is an important quality for the morphogen gradient.
\section{Discussion \& Conclusion}
We studied the formation of a stationary gradient for a nonlinear degradation of morphogens with non-Markovian transport.
In particular, this interaction leads to the phenomenon of a \textit{degradation enhanced diffusion} in the long-time limit.
We see that a increase in the rate of degradation actually leads to an increase in diffusion.
Additionally we have shown that the stationary profile is no longer of exponential type, rather, it is of power law type.
The shape of the tails \eqref{eq:sub_tails} is determined by the anomalous exponent $ \mu $.
The stationary solution as $ x \to \infty $ is actually independent from the effects of the production rate entirely.
It is well known that the importance of the power law profile is due to its robustness to fluctuations in the production rate of morphogens, and also to other environmental effects.

We made a connection between the non-Markovian subdiffusive model with nonlinear reaction and the Markovian nonlinear reaction-diffusion equations \cite{Boon2012}.
When, in the fractional formulation, we assume a constant value of anomalous exponent $ \mu $ and a power-law ansatz for the reactive term; the steady state reaction-subdiffusion equation takes the same form as that obtained from the non-linear reaction-diffusion equation.
The essential point of our paper is that that we have not just studied nonlinear diffusion, but derived the nonlinear dependence of diffusion on the nonlinear reaction in the long time limit.
In fact, we should note that this result can be extended to the general non-Markovian transport process.
This is a subject for future work.

\textit{Acknowledgements.} 
Sergei Fedotov gratefully acknowledges the support of the EPSRC Grant EP/J019526/1.
The authors wish to thank Peter Straka and Hugues Berry for interesting discussions.
\bibliographystyle{apsrev4-1}
\bibliography{ibb}

\begin{thebibliography}{32}%
\makeatletter
\providecommand \@ifxundefined [1]{%
 \@ifx{#1\undefined}
}%
\providecommand \@ifnum [1]{%
 \ifnum #1\expandafter \@firstoftwo
 \else \expandafter \@secondoftwo
 \fi
}%
\providecommand \@ifx [1]{%
 \ifx #1\expandafter \@firstoftwo
 \else \expandafter \@secondoftwo
 \fi
}%
\providecommand \natexlab [1]{#1}%
\providecommand \enquote  [1]{``#1''}%
\providecommand \bibnamefont  [1]{#1}%
\providecommand \bibfnamefont [1]{#1}%
\providecommand \citenamefont [1]{#1}%
\providecommand \href@noop [0]{\@secondoftwo}%
\providecommand \href [0]{\begingroup \@sanitize@url \@href}%
\providecommand \@href[1]{\@@startlink{#1}\@@href}%
\providecommand \@@href[1]{\endgroup#1\@@endlink}%
\providecommand \@sanitize@url [0]{\catcode `\\12\catcode `\$12\catcode
  `\&12\catcode `\#12\catcode `\^12\catcode `\_12\catcode `\%12\relax}%
\providecommand \@@startlink[1]{}%
\providecommand \@@endlink[0]{}%
\providecommand \url  [0]{\begingroup\@sanitize@url \@url }%
\providecommand \@url [1]{\endgroup\@href {#1}{\urlprefix }}%
\providecommand \urlprefix  [0]{URL }%
\providecommand \Eprint [0]{\href }%
\providecommand \doibase [0]{http://dx.doi.org/}%
\providecommand \selectlanguage [0]{\@gobble}%
\providecommand \bibinfo  [0]{\@secondoftwo}%
\providecommand \bibfield  [0]{\@secondoftwo}%
\providecommand \translation [1]{[#1]}%
\providecommand \BibitemOpen [0]{}%
\providecommand \bibitemStop [0]{}%
\providecommand \bibitemNoStop [0]{.\EOS\space}%
\providecommand \EOS [0]{\spacefactor3000\relax}%
\providecommand \BibitemShut  [1]{\csname bibitem#1\endcsname}%
\let\auto@bib@innerbib\@empty
\bibitem [{\citenamefont {Rogers}\ and\ \citenamefont
  {Schier}(2011)}]{Rogers2011}%
  \BibitemOpen
  \bibfield  {author} {\bibinfo {author} {\bibfnamefont {K.~W.}\ \bibnamefont
  {Rogers}}\ and\ \bibinfo {author} {\bibfnamefont {A.~F.}\ \bibnamefont
  {Schier}},\ }\href {\doibase 10.1146/annurev-cellbio-092910-154148}
  {\bibfield  {journal} {\bibinfo  {journal} {Annual Review of Cell and
  Developmental Biology}\ }\textbf {\bibinfo {volume} {27}},\ \bibinfo {pages}
  {377} (\bibinfo {year} {2011})}\BibitemShut {NoStop}%
\bibitem [{\citenamefont {Entchev}\ \emph {et~al.}(2000)\citenamefont
  {Entchev}, \citenamefont {Schwabedissen},\ and\ \citenamefont
  {Gonz\'{a}lez-Gait\'{a}n}}]{Entchev2000}%
  \BibitemOpen
  \bibfield  {author} {\bibinfo {author} {\bibfnamefont {E.~V.}\ \bibnamefont
  {Entchev}}, \bibinfo {author} {\bibfnamefont {A.}~\bibnamefont
  {Schwabedissen}}, \ and\ \bibinfo {author} {\bibfnamefont {M.}~\bibnamefont
  {Gonz\'{a}lez-Gait\'{a}n}},\ }\href
  {http://www.ncbi.nlm.nih.gov/pubmed/11136982} {\bibfield  {journal} {\bibinfo
   {journal} {Cell}\ }\textbf {\bibinfo {volume} {103}},\ \bibinfo {pages}
  {981} (\bibinfo {year} {2000})}\BibitemShut {NoStop}%
\bibitem [{\citenamefont {Barkai}\ and\ \citenamefont
  {Shilo}(2009)}]{Barkai2009}%
  \BibitemOpen
  \bibfield  {author} {\bibinfo {author} {\bibfnamefont {N.}~\bibnamefont
  {Barkai}}\ and\ \bibinfo {author} {\bibfnamefont {B.-Z.}\ \bibnamefont
  {Shilo}},\ }\href
  {http://cshperspectives.cshlp.org/content/1/5/a001990.abstract} {\bibfield
  {journal} {\bibinfo  {journal} {Cold Spring Harbor Perspectives in Biology}\
  }\textbf {\bibinfo {volume} {1}} (\bibinfo {year} {2009})}\BibitemShut
  {NoStop}%
\bibitem [{\citenamefont {Ashe}\ and\ \citenamefont
  {Briscoe}(2006)}]{Ashe2006}%
  \BibitemOpen
  \bibfield  {author} {\bibinfo {author} {\bibfnamefont {H.~L.}\ \bibnamefont
  {Ashe}}\ and\ \bibinfo {author} {\bibfnamefont {J.}~\bibnamefont {Briscoe}},\
  }\href {\doibase 10.1242/dev.02238} {\bibfield  {journal} {\bibinfo
  {journal} {Development}\ }\textbf {\bibinfo {volume} {133}},\ \bibinfo
  {pages} {385} (\bibinfo {year} {2006})}\BibitemShut {NoStop}%
\bibitem [{\citenamefont {Eldar}\ \emph {et~al.}(2002)\citenamefont {Eldar},
  \citenamefont {Dorfman}, \citenamefont {Weiss},\ and\ \citenamefont
  {Ashe}}]{Eldar2002}%
  \BibitemOpen
  \bibfield  {author} {\bibinfo {author} {\bibfnamefont {A.}~\bibnamefont
  {Eldar}}, \bibinfo {author} {\bibfnamefont {R.}~\bibnamefont {Dorfman}},
  \bibinfo {author} {\bibfnamefont {D.}~\bibnamefont {Weiss}}, \ and\ \bibinfo
  {author} {\bibfnamefont {H.}~\bibnamefont {Ashe}},\ }\href
  {http://www.nature.com/nature/journal/v419/n6904/abs/nature01061.html}
  {\bibfield  {journal} {\bibinfo  {journal} {Nature}\ }\textbf {\bibinfo
  {volume} {419}} (\bibinfo {year} {2002})}\BibitemShut {NoStop}%
\bibitem [{\citenamefont {Bressloff}\ and\ \citenamefont
  {Newby}(2013)}]{Bressloff2013}%
  \BibitemOpen
  \bibfield  {author} {\bibinfo {author} {\bibfnamefont {P.~C.}\ \bibnamefont
  {Bressloff}}\ and\ \bibinfo {author} {\bibfnamefont {J.~M.}\ \bibnamefont
  {Newby}},\ }\href {\doibase 10.1103/RevModPhys.85.135} {\bibfield  {journal}
  {\bibinfo  {journal} {Rev. Mod. Phys.}\ }\textbf {\bibinfo {volume} {85}},\
  \bibinfo {pages} {135} (\bibinfo {year} {2013})}\BibitemShut {NoStop}%
\bibitem [{\citenamefont {Bollenbach}\ \emph {et~al.}(2005)\citenamefont
  {Bollenbach}, \citenamefont {Kruse}, \citenamefont {Pantazis}, \citenamefont
  {Gonz\'{a}lez-Gait\'{a}n},\ and\ \citenamefont
  {J\"{u}licher}}]{Bollenbach2005}%
  \BibitemOpen
  \bibfield  {author} {\bibinfo {author} {\bibfnamefont {T.}~\bibnamefont
  {Bollenbach}}, \bibinfo {author} {\bibfnamefont {K.}~\bibnamefont {Kruse}},
  \bibinfo {author} {\bibfnamefont {P.}~\bibnamefont {Pantazis}}, \bibinfo
  {author} {\bibfnamefont {M.}~\bibnamefont {Gonz\'{a}lez-Gait\'{a}n}}, \ and\
  \bibinfo {author} {\bibfnamefont {F.}~\bibnamefont {J\"{u}licher}},\ }\href
  {\doibase 10.1103/PhysRevLett.94.018103} {\bibfield  {journal} {\bibinfo
  {journal} {Phys. Rev. Lett.}\ }\textbf {\bibinfo {volume} {94}},\ \bibinfo
  {pages} {018103} (\bibinfo {year} {2005})}\BibitemShut {NoStop}%
\bibitem [{\citenamefont {Kicheva}\ \emph {et~al.}(2007)\citenamefont
  {Kicheva}, \citenamefont {Pantazis}, \citenamefont {Bollenbach},
  \citenamefont {Kalaidzidis}, \citenamefont {Bittig}, \citenamefont
  {J\"{u}licher},\ and\ \citenamefont {Gonz\'{a}lez-Gait\'{a}n}}]{Kicheva2007}%
  \BibitemOpen
  \bibfield  {author} {\bibinfo {author} {\bibfnamefont {A.}~\bibnamefont
  {Kicheva}}, \bibinfo {author} {\bibfnamefont {P.}~\bibnamefont {Pantazis}},
  \bibinfo {author} {\bibfnamefont {T.}~\bibnamefont {Bollenbach}}, \bibinfo
  {author} {\bibfnamefont {Y.}~\bibnamefont {Kalaidzidis}}, \bibinfo {author}
  {\bibfnamefont {T.}~\bibnamefont {Bittig}}, \bibinfo {author} {\bibfnamefont
  {F.}~\bibnamefont {J\"{u}licher}}, \ and\ \bibinfo {author} {\bibfnamefont
  {M.}~\bibnamefont {Gonz\'{a}lez-Gait\'{a}n}},\ }\href {\doibase
  10.1126/science.1135774} {\bibfield  {journal} {\bibinfo  {journal}
  {Science}\ }\textbf {\bibinfo {volume} {315}},\ \bibinfo {pages} {521}
  (\bibinfo {year} {2007})}\BibitemShut {NoStop}%
\bibitem [{\citenamefont {Eldar}\ \emph
  {et~al.}(2003{\natexlab{a}})\citenamefont {Eldar}, \citenamefont {Rosin},
  \citenamefont {Shilo},\ and\ \citenamefont {Barkai}}]{Eldar2003}%
  \BibitemOpen
  \bibfield  {author} {\bibinfo {author} {\bibfnamefont {A.}~\bibnamefont
  {Eldar}}, \bibinfo {author} {\bibfnamefont {D.}~\bibnamefont {Rosin}},
  \bibinfo {author} {\bibfnamefont {B.-Z.}\ \bibnamefont {Shilo}}, \ and\
  \bibinfo {author} {\bibfnamefont {N.}~\bibnamefont {Barkai}},\ }\href
  {http://www.ncbi.nlm.nih.gov/pubmed/14536064} {\bibfield  {journal} {\bibinfo
   {journal} {Developmental Cell}\ }\textbf {\bibinfo {volume} {5}},\ \bibinfo
  {pages} {635} (\bibinfo {year} {2003}{\natexlab{a}})}\BibitemShut {NoStop}%
\bibitem [{\citenamefont {Han}\ \emph {et~al.}(2005)\citenamefont {Han},
  \citenamefont {Yan}, \citenamefont {Belenkaya},\ and\ \citenamefont
  {Lin}}]{Han2005}%
  \BibitemOpen
  \bibfield  {author} {\bibinfo {author} {\bibfnamefont {C.}~\bibnamefont
  {Han}}, \bibinfo {author} {\bibfnamefont {D.}~\bibnamefont {Yan}}, \bibinfo
  {author} {\bibfnamefont {T.~Y.}\ \bibnamefont {Belenkaya}}, \ and\ \bibinfo
  {author} {\bibfnamefont {X.}~\bibnamefont {Lin}},\ }\href {\doibase
  10.1242/dev.01636} {\bibfield  {journal} {\bibinfo  {journal} {Development}\
  }\textbf {\bibinfo {volume} {132}},\ \bibinfo {pages} {667} (\bibinfo {year}
  {2005})}\BibitemShut {NoStop}%
\bibitem [{\citenamefont {Chen}\ and\ \citenamefont {Struhl}(1996)}]{Chen1996}%
  \BibitemOpen
  \bibfield  {author} {\bibinfo {author} {\bibfnamefont {Y.}~\bibnamefont
  {Chen}}\ and\ \bibinfo {author} {\bibfnamefont {G.}~\bibnamefont {Struhl}},\
  }\href {http://www.ncbi.nlm.nih.gov/pubmed/8898207} {\bibfield  {journal}
  {\bibinfo  {journal} {Cell}\ }\textbf {\bibinfo {volume} {87}},\ \bibinfo
  {pages} {553} (\bibinfo {year} {1996})}\BibitemShut {NoStop}%
\bibitem [{\citenamefont {Gordon}\ \emph {et~al.}(2011)\citenamefont {Gordon},
  \citenamefont {Sample}, \citenamefont {Berezhkovskii}, \citenamefont
  {Muratov},\ and\ \citenamefont {Shvartsman}}]{Gordon2011a}%
  \BibitemOpen
  \bibfield  {author} {\bibinfo {author} {\bibfnamefont {P.~V.}\ \bibnamefont
  {Gordon}}, \bibinfo {author} {\bibfnamefont {C.}~\bibnamefont {Sample}},
  \bibinfo {author} {\bibfnamefont {A.~M.}\ \bibnamefont {Berezhkovskii}},
  \bibinfo {author} {\bibfnamefont {C.~B.}\ \bibnamefont {Muratov}}, \ and\
  \bibinfo {author} {\bibfnamefont {S.~Y.}\ \bibnamefont {Shvartsman}},\ }\href
  {\doibase 10.1073/pnas.1019245108} {\bibfield  {journal} {\bibinfo  {journal}
  {PNAS}\ }\textbf {\bibinfo {volume} {108}},\ \bibinfo {pages} {6157}
  (\bibinfo {year} {2011})}\BibitemShut {NoStop}%
\bibitem [{\citenamefont {Muratov}\ \emph {et~al.}(2011)\citenamefont
  {Muratov}, \citenamefont {Gordon},\ and\ \citenamefont
  {Shvartsman}}]{Muratov2011}%
  \BibitemOpen
  \bibfield  {author} {\bibinfo {author} {\bibfnamefont {C.~B.}\ \bibnamefont
  {Muratov}}, \bibinfo {author} {\bibfnamefont {P.~V.}\ \bibnamefont {Gordon}},
  \ and\ \bibinfo {author} {\bibfnamefont {S.~Y.}\ \bibnamefont {Shvartsman}},\
  }\href@noop {} {\bibfield  {journal} {\bibinfo  {journal} {Physical Review
  E}\ }\textbf {\bibinfo {volume} {84}},\ \bibinfo {pages} {041916} (\bibinfo
  {year} {2011})}\BibitemShut {NoStop}%
\bibitem [{\citenamefont {Hornung}\ \emph {et~al.}(2005)\citenamefont
  {Hornung}, \citenamefont {Berkowitz},\ and\ \citenamefont
  {Barkai}}]{Hornung}%
  \BibitemOpen
  \bibfield  {author} {\bibinfo {author} {\bibfnamefont {G.}~\bibnamefont
  {Hornung}}, \bibinfo {author} {\bibfnamefont {B.}~\bibnamefont {Berkowitz}},
  \ and\ \bibinfo {author} {\bibfnamefont {N.}~\bibnamefont {Barkai}},\ }\href
  {\doibase 10.1103/PhysRevE.72.041916} {\bibfield  {journal} {\bibinfo
  {journal} {Phys. Rev. E}\ }\textbf {\bibinfo {volume} {72}},\ \bibinfo
  {pages} {041916} (\bibinfo {year} {2005})}\BibitemShut {NoStop}%
\bibitem [{\citenamefont {Weiss}\ \emph {et~al.}(2004)\citenamefont {Weiss},
  \citenamefont {Elsner}, \citenamefont {Kartberg},\ and\ \citenamefont
  {Nilsson}}]{Weiss2004}%
  \BibitemOpen
  \bibfield  {author} {\bibinfo {author} {\bibfnamefont {M.}~\bibnamefont
  {Weiss}}, \bibinfo {author} {\bibfnamefont {M.}~\bibnamefont {Elsner}},
  \bibinfo {author} {\bibfnamefont {F.}~\bibnamefont {Kartberg}}, \ and\
  \bibinfo {author} {\bibfnamefont {T.}~\bibnamefont {Nilsson}},\ }\href
  {\doibase 10.1529/biophysj.104.044263} {\bibfield  {journal} {\bibinfo
  {journal} {Biophysical Journal}\ }\textbf {\bibinfo {volume} {87}},\ \bibinfo
  {pages} {3518} (\bibinfo {year} {2004})}\BibitemShut {NoStop}%
\bibitem [{\citenamefont {Golding}\ and\ \citenamefont
  {Cox}(2006)}]{Golding2006}%
  \BibitemOpen
  \bibfield  {author} {\bibinfo {author} {\bibfnamefont {I.}~\bibnamefont
  {Golding}}\ and\ \bibinfo {author} {\bibfnamefont {E.~C.}\ \bibnamefont
  {Cox}},\ }\href {\doibase 10.1103/PhysRevLett.96.098102} {\bibfield
  {journal} {\bibinfo  {journal} {Phys. Rev. Lett.}\ }\textbf {\bibinfo
  {volume} {96}},\ \bibinfo {pages} {098102} (\bibinfo {year}
  {2006})}\BibitemShut {NoStop}%
\bibitem [{\citenamefont {Saxton}(2001)}]{Saxton2001}%
  \BibitemOpen
  \bibfield  {author} {\bibinfo {author} {\bibfnamefont {M.~J.}\ \bibnamefont
  {Saxton}},\ }\href {\doibase 10.1016/S0006-3495(01)75870-5} {\bibfield
  {journal} {\bibinfo  {journal} {Biophysical Journal}\ }\textbf {\bibinfo
  {volume} {81}},\ \bibinfo {pages} {2226} (\bibinfo {year}
  {2001})}\BibitemShut {NoStop}%
\bibitem [{\citenamefont {Min}\ \emph {et~al.}(2005)\citenamefont {Min},
  \citenamefont {Luo}, \citenamefont {Cherayil}, \citenamefont {Kou},\ and\
  \citenamefont {Xie}}]{Min2005}%
  \BibitemOpen
  \bibfield  {author} {\bibinfo {author} {\bibfnamefont {W.}~\bibnamefont
  {Min}}, \bibinfo {author} {\bibfnamefont {G.}~\bibnamefont {Luo}}, \bibinfo
  {author} {\bibfnamefont {B.~J.}\ \bibnamefont {Cherayil}}, \bibinfo {author}
  {\bibfnamefont {S.~C.}\ \bibnamefont {Kou}}, \ and\ \bibinfo {author}
  {\bibfnamefont {X.~S.}\ \bibnamefont {Xie}},\ }\href {\doibase
  10.1103/PhysRevLett.94.198302} {\bibfield  {journal} {\bibinfo  {journal}
  {Phys. Rev. Lett.}\ }\textbf {\bibinfo {volume} {94}},\ \bibinfo {pages}
  {198302} (\bibinfo {year} {2005})}\BibitemShut {NoStop}%
\bibitem [{\citenamefont {Kruse}\ and\ \citenamefont
  {Iomin}(2008)}]{Kruse2008}%
  \BibitemOpen
  \bibfield  {author} {\bibinfo {author} {\bibfnamefont {K.}~\bibnamefont
  {Kruse}}\ and\ \bibinfo {author} {\bibfnamefont {A.}~\bibnamefont {Iomin}},\
  }\href {\doibase 10.1088/1367-2630/10/2/023019} {\bibfield  {journal}
  {\bibinfo  {journal} {New Journal of Physics}\ }\textbf {\bibinfo {volume}
  {10}},\ \bibinfo {pages} {023019} (\bibinfo {year} {2008})}\BibitemShut
  {NoStop}%
\bibitem [{\citenamefont {Yuste}\ \emph {et~al.}(2010)\citenamefont {Yuste},
  \citenamefont {Abad},\ and\ \citenamefont {Lindenberg}}]{Yuste2010}%
  \BibitemOpen
  \bibfield  {author} {\bibinfo {author} {\bibfnamefont {S.~B.}\ \bibnamefont
  {Yuste}}, \bibinfo {author} {\bibfnamefont {E.}~\bibnamefont {Abad}}, \ and\
  \bibinfo {author} {\bibfnamefont {K.}~\bibnamefont {Lindenberg}},\ }\href
  {\doibase 10.1103/PhysRevE.82.061123} {\bibfield  {journal} {\bibinfo
  {journal} {Phys. Rev. E}\ }\textbf {\bibinfo {volume} {82}},\ \bibinfo
  {pages} {061123} (\bibinfo {year} {2010})}\BibitemShut {NoStop}%
\bibitem [{\citenamefont {Boon}\ \emph {et~al.}(2012)\citenamefont {Boon},
  \citenamefont {Lutsko},\ and\ \citenamefont {Lutsko}}]{Boon2012}%
  \BibitemOpen
  \bibfield  {author} {\bibinfo {author} {\bibfnamefont {J.~P.}\ \bibnamefont
  {Boon}}, \bibinfo {author} {\bibfnamefont {J.~F.}\ \bibnamefont {Lutsko}}, \
  and\ \bibinfo {author} {\bibfnamefont {C.}~\bibnamefont {Lutsko}},\ }\href
  {\doibase 10.1103/PhysRevE.85.021126} {\bibfield  {journal} {\bibinfo
  {journal} {Phys. Rev. E}\ }\textbf {\bibinfo {volume} {85}},\ \bibinfo
  {pages} {021126} (\bibinfo {year} {2012})}\BibitemShut {NoStop}%
\bibitem [{\citenamefont {Fedotov}\ and\ \citenamefont
  {Falconer}(2013)}]{Fedotov2013}%
  \BibitemOpen
  \bibfield  {author} {\bibinfo {author} {\bibfnamefont {S.}~\bibnamefont
  {Fedotov}}\ and\ \bibinfo {author} {\bibfnamefont {S.}~\bibnamefont
  {Falconer}},\ }\href {\doibase 10.1103/PhysRevE.87.052139} {\bibfield
  {journal} {\bibinfo  {journal} {Phys. Rev. E}\ }\textbf {\bibinfo {volume}
  {87}},\ \bibinfo {pages} {052139} (\bibinfo {year} {2013})}\BibitemShut
  {NoStop}%
\bibitem [{\citenamefont {Kolomeisky}(2011)}]{Anatoly}%
  \BibitemOpen
  \bibfield  {author} {\bibinfo {author} {\bibfnamefont {A.~B.}\ \bibnamefont
  {Kolomeisky}},\ }\href@noop {} {\bibfield  {journal} {\bibinfo  {journal}
  {The Journal of Physical Chemistry Letters}\ }\textbf {\bibinfo {volume}
  {2}},\ \bibinfo {pages} {1502} (\bibinfo {year} {2011})}\BibitemShut
  {NoStop}%
\bibitem [{\citenamefont {Hilfer}\ and\ \citenamefont {Anton}(1995)}]{Hilfer}%
  \BibitemOpen
  \bibfield  {author} {\bibinfo {author} {\bibfnamefont {R.}~\bibnamefont
  {Hilfer}}\ and\ \bibinfo {author} {\bibfnamefont {L.}~\bibnamefont {Anton}},\
  }\href {\doibase 10.1103/PhysRevE.51.R848} {\bibfield  {journal} {\bibinfo
  {journal} {Phys. Rev. E}\ }\textbf {\bibinfo {volume} {51}},\ \bibinfo
  {pages} {R848} (\bibinfo {year} {1995})}\BibitemShut {NoStop}%
\bibitem [{\citenamefont {Santamaria}\ \emph {et~al.}(2011)\citenamefont
  {Santamaria}, \citenamefont {Wils}, \citenamefont {{De Schutter}},\ and\
  \citenamefont {Augustine}}]{Santamaria2011}%
  \BibitemOpen
  \bibfield  {author} {\bibinfo {author} {\bibfnamefont {F.}~\bibnamefont
  {Santamaria}}, \bibinfo {author} {\bibfnamefont {S.}~\bibnamefont {Wils}},
  \bibinfo {author} {\bibfnamefont {E.}~\bibnamefont {{De Schutter}}}, \ and\
  \bibinfo {author} {\bibfnamefont {G.~J.}\ \bibnamefont {Augustine}},\ }\href
  {\doibase 10.1111/j.1460-9568.2011.07785.x} {\bibfield  {journal} {\bibinfo
  {journal} {European Journal of Neuroscience}\ }\textbf {\bibinfo {volume}
  {34}},\ \bibinfo {pages} {561} (\bibinfo {year} {2011})}\BibitemShut
  {NoStop}%
\bibitem [{\citenamefont {Yadav}\ and\ \citenamefont
  {Horsthemke}(2006)}]{Yadav2006}%
  \BibitemOpen
  \bibfield  {author} {\bibinfo {author} {\bibfnamefont {A.}~\bibnamefont
  {Yadav}}\ and\ \bibinfo {author} {\bibfnamefont {W.}~\bibnamefont
  {Horsthemke}},\ }\href {\doibase 10.1103/PhysRevE.74.066118} {\bibfield
  {journal} {\bibinfo  {journal} {Phys. Rev. E}\ }\textbf {\bibinfo {volume}
  {74}},\ \bibinfo {pages} {066118} (\bibinfo {year} {2006})}\BibitemShut
  {NoStop}%
\bibitem [{\citenamefont {M\'{e}ndez}\ \emph {et~al.}(2010)\citenamefont
  {M\'{e}ndez}, \citenamefont {Fedotov},\ and\ \citenamefont
  {Horsthemke}}]{mendez2010reaction}%
  \BibitemOpen
  \bibfield  {author} {\bibinfo {author} {\bibfnamefont {V.}~\bibnamefont
  {M\'{e}ndez}}, \bibinfo {author} {\bibfnamefont {S.}~\bibnamefont {Fedotov}},
  \ and\ \bibinfo {author} {\bibfnamefont {W.}~\bibnamefont {Horsthemke}},\
  }\href {http://books.google.co.uk/books?id=RxjoTmjAwZYC} {\emph {\bibinfo
  {title} {Reaction-Transport Systems: Mesoscopic Foundations,\ Fronts, and
  Spatial Instabilities}}},\ Springer Series in Synergetics\ (\bibinfo
  {publisher} {Springer},\ \bibinfo {year} {2010})\BibitemShut {NoStop}%
\bibitem [{\citenamefont {Fedotov}\ and\ \citenamefont
  {Falconer}(2012)}]{Fedotov2012}%
  \BibitemOpen
  \bibfield  {author} {\bibinfo {author} {\bibfnamefont {S.}~\bibnamefont
  {Fedotov}}\ and\ \bibinfo {author} {\bibfnamefont {S.}~\bibnamefont
  {Falconer}},\ }\href {\doibase 10.1103/PhysRevE.85.031132} {\bibfield
  {journal} {\bibinfo  {journal} {Phys. Rev. E}\ }\textbf {\bibinfo {volume}
  {85}},\ \bibinfo {pages} {031132} (\bibinfo {year} {2012})}\BibitemShut
  {NoStop}%
\bibitem [{\citenamefont {Fedotov}(2011)}]{Fedotov2011}%
  \BibitemOpen
  \bibfield  {author} {\bibinfo {author} {\bibfnamefont {S.}~\bibnamefont
  {Fedotov}},\ }\href {\doibase 10.1103/PhysRevE.83.021110} {\bibfield
  {journal} {\bibinfo  {journal} {Phys. Rev. E}\ }\textbf {\bibinfo {volume}
  {83}},\ \bibinfo {pages} {021110} (\bibinfo {year} {2011})}\BibitemShut
  {NoStop}%
\bibitem [{\citenamefont {Fedotov}(2013)}]{fedotov2013nonlinear}%
  \BibitemOpen
  \bibfield  {author} {\bibinfo {author} {\bibfnamefont {S.}~\bibnamefont
  {Fedotov}},\ }\href {\doibase 10.1103/PhysRevE.88.032104} {\bibfield
  {journal} {\bibinfo  {journal} {Phys. Rev. E}\ }\textbf {\bibinfo {volume}
  {88}},\ \bibinfo {pages} {032104} (\bibinfo {year} {2013})}\BibitemShut
  {NoStop}%
\bibitem [{\citenamefont {Froemberg}\ and\ \citenamefont
  {Sokolov}(2008)}]{Froemberg2008}%
  \BibitemOpen
  \bibfield  {author} {\bibinfo {author} {\bibfnamefont {D.}~\bibnamefont
  {Froemberg}}\ and\ \bibinfo {author} {\bibfnamefont {I.~M.}\ \bibnamefont
  {Sokolov}},\ }\href {\doibase 10.1103/PhysRevLett.100.108304} {\bibfield
  {journal} {\bibinfo  {journal} {Phys. Rev. Lett.}\ }\textbf {\bibinfo
  {volume} {100}},\ \bibinfo {pages} {108304} (\bibinfo {year}
  {2008})}\BibitemShut {NoStop}%
\bibitem [{\citenamefont {Eldar}\ \emph
  {et~al.}(2003{\natexlab{b}})\citenamefont {Eldar}, \citenamefont {Rosin},
  \citenamefont {Shilo},\ and\ \citenamefont {Barkai}}]{Eldar2003b}%
  \BibitemOpen
  \bibfield  {author} {\bibinfo {author} {\bibfnamefont {A.}~\bibnamefont
  {Eldar}}, \bibinfo {author} {\bibfnamefont {D.}~\bibnamefont {Rosin}},
  \bibinfo {author} {\bibfnamefont {B.-Z.}\ \bibnamefont {Shilo}}, \ and\
  \bibinfo {author} {\bibfnamefont {N.}~\bibnamefont {Barkai}},\ }\href
  {\doibase http://dx.doi.org/10.1016/S1534-5807(03)00292-2} {\bibfield
  {journal} {\bibinfo  {journal} {Developmental Cell}\ }\textbf {\bibinfo
  {volume} {5}},\ \bibinfo {pages} {635 } (\bibinfo {year}
  {2003}{\natexlab{b}})}\BibitemShut {NoStop}%
\end{thebibliography}%

\end{document}